**Optical polarization of nuclear spins in silicon carbide**

Abram L. Falk[1,2], Paul V. Klimov[1,3], Viktor Ivády[4,5], Krisztián Szász[4,6], David J. Christle[1,3], William F. Koehl[1], Ádám Gali[4,7], and David D. Awschalom[1,*]

1. Institute for Molecular Engineering, University of Chicago, Chicago, IL 60637, USA
2. IBM T. J. Watson Research Center, 1101 Kitchawan Rd., Yorktown Heights, NY 10598, USA
3. Center for Spintronics and Quantum Computation, University of California, Santa Barbara, Santa Barbara, CA 93106, USA
4. Institute for Solid State Physics and Optics, Wigner Research Centre for Physics, Hungarian Academy of Sciences, Budapest, Hungary
5. Department of Physics, Chemistry, and Biology, Linköping University, Sweden
6. Institute of Physics, Loránd Eötvös University, Hungary
7. Department of Atomic Physics, Budapest University of Technology and Economics, Budapest, Hungary
* Email: awsch@uchicago.edu

**Abstract:**

We demonstrate optically pumped dynamic nuclear polarization of $^{29}$Si nuclear spins that are strongly coupled to paramagnetic color centers in 4*H*- and 6*H*-SiC. The 99 ± 1% degree of polarization at room temperature corresponds to an effective nuclear temperature of 5 μK. By combining *ab initio* theory with the experimental identification of the color centers' optically excited states, we quantitatively model how the polarization derives from hyperfine-mediated level anticrossings. These results lay a foundation for SiC-based quantum memories, nuclear gyroscopes, and hyperpolarized probes for magnetic resonance imaging.

**Main text:**

Silicon carbide is a promising material for quantum electronics at the wafer scale. It is both amenable to sophisticated device processing [1] and a host for several types of vacancy-related paramagnetic color centers with remarkable attributes [2-23]. Much like the diamond nitrogen-vacancy center [24, 25], these color centers localize electronic states that can exhibit millisecond-long spin coherence times [18], single-spin addressability through confocal optically detected magnetic resonance (ODMR) [18, 19], and ODMR persistence up to room temperature [10, 11, 14, 19]. Although fluctuating nuclear spins are a principal source of electronic spin decoherence [26], their presence is not purely detrimental. If polarized and controlled, nuclear spins in SiC would be a technologically important resource.

In this Letter, we show that near-infrared light can nearly completely polarize populations of $^{29}$Si nuclear spins in SiC. This process is based on dynamic nuclear polarization (DNP) [27, 28]: the optically pumped polarization of of electron spins bound to either neutral divacancy [4, 5, 8, 10, 14] or PL6 [10, 14, 16] color centers is transferred to proximate nuclei via the hyperfine interaction. Realizing DNP in SiC is experimentally straightforward, requiring only non-resonant optical illumination and a small external magnetic field (300–500 G), which tunes color-center ensembles to either their ground-state (GS) or excited-state (ES) level anticrossings (the GSLAC and ESLAC, respectively). Optically pumping room-temperature crystals has previously led to DNP in napthalene [29], diamond [30-35], and GaNAs [36]. Our results show that strong room-temperature DNP can be driven in a material that plays a leading role in the semiconductor industry.

We find that SiC color centers can mediate a high degree (>85%) of ESLAC-derived nuclear polarization from 5–298 K, a surprisingly broad temperature range. This robust DNP could be applied to initialize quantum memories in quantum-communication technologies, especially since the color centers are telecom-range emitters, with narrow optical linewidths at low temperatures [16, 21, 37]. Other applications of DNP, including solid-state nuclear gyroscopes [38, 39] and entanglement-enhanced metrological devices [40], can employ SiC's long nuclear spin-lattice relaxation times [27, 41] and the significant density ($10^{16}$ cm$^{-3}$ [42]) of defect-coupled nuclei that we can polarize.

The divacancy defect in SiC is a silicon vacancy ($V_{Si}$) adjacent to a carbon vacancy ($V_C$) (Fig. 1a). Because vacancies in hexagonal (*h*) and quasicubic (*k*) lattice sites are inequivalent, divacancies also have several inequivalent forms. Those aligned to the crystal's *c* axis, which we study in this work, are the *hh* and *kk* divacancies in 4*H*-SiC [4, 5], and the *hh*, *k₁k₁* and *k₂k₂* divacancies in 6*H*-SiC [14, 43, 44]. The physical structure of the *c*-axis-oriented PL6 defect in 4*H*-SiC [10, 14, 16] is currently undetermined, but a close relationship to the neutral divacancies is indicated by its similar optical and spin resonances [10], similar radiative lifetimes [16], and similar hyperfine interaction strengths and degeneracies, which we measure here (Table 1). In their GS, these defects are spin triplets ($S = 1$) with the Hamiltonian:

$$H_{GS} = \mu_B \mathbf{B} \cdot \mathbf{g}_{GS} \cdot \mathbf{S} + D_{GS} S_z^2 + \sum_j \gamma_j \mathbf{B} \cdot \mathbf{I}_j + \mathbf{S} \cdot \mathbf{A}_{j,GS} \cdot \mathbf{I}_j$$

(1)

where $\mathbf{g}_{GS}$ is the electronic *g*-tensor, $\mu_B$ the Bohr magneton, $\mathbf{B}$ is the external magnetic field, $D_{GS}$ is the electronic zero-field splitting parameter, $\mathbf{S}$ is the vector of electron spin operators, and $\mathbf{A}_{j,GS}$ the hyperfine tensor of the $j^{th}$ nearby nucleus with spin $\mathbf{I}_j$ and gyromagnetic ratio $\gamma_j$. The four terms in Eq. (1) represent the electron's Zeeman effect, the electronic crystal-field splitting, the nuclear Zeeman effect, and the hyperfine interaction between electronic and nuclear spins. At elevated temperatures, the ES has a similar

Hamiltonian to that of the GS but with different parameter values: $g_{ES}$ substituting for $g_{GS}$, $D_{ES}$ substituting for $D_{GS}$, and $\mathbf{A}_{j,ES}$ substituting for $\mathbf{A}_{j,GS}$.

Silicon's dominant isotope is spin-0 $^{28}$Si, but the spin-1/2 isotope $^{29}$Si also has a fairly high natural abundance of 4.7%. We denote the state of a hyperfine-coupled spin pair as $|m_S, m_I\rangle$, where $m_S \in \{-1, 0, 1\}$ is the electronic spin state and $m_I \in \{\uparrow, \downarrow\}$ is the $^{29}$Si nuclear spin state. Before any optical pumping, the spin pairs are a statistical mixture of all the $|m_S, m_I\rangle$ states.

Optical illumination polarizes the color centers' electronic spins into the $m_S = 0$ sublevel, a consequence of spin-dependent coupling to an intersystem crossing during the optical cycle [8, 14, 18, 21]. We infer this sign of optically pumped electron polarization using double electron-electron resonance [14, 42]. On its own, optical cycling does not polarize nuclear spins and results in equal populations of $|0, \downarrow\rangle$ and $|0, \uparrow\rangle$ states. However, when the defects' spin sublevels are tuned to be near a level anticrossing (either the ESLAC or GSLAC), the hyperfine interaction mixes $|0, \downarrow\rangle$ with $|-1, \uparrow\rangle$. This mixing causes $|0, \downarrow\rangle$ to be a non-stationary state of the ES- or GS-spin Hamiltonian and to partially evolve into $|-1, \uparrow\rangle$ during each optical cycle. Further optical cycles then reorient the electronic spins, polarizing $|-1, \uparrow\rangle$ states into $|0, \uparrow\rangle$. Meanwhile, conservation of angular momentum prevents $|0, \uparrow\rangle$ from mixing with $|-1, \downarrow\rangle$. Together, these processes can efficiently polarize arbitrary $|m_S, m_I\rangle$ states into $|0, \uparrow\rangle$ (Figs. 1b-d).

Our 4*H*-SiC wafer (purchased from Cree, Inc.) has vacancy complexes intentionally incorporated during crystal growth [10]. In our 6*H*-SiC wafer (purchased from II-VI, Inc.), we generate divacancies by bombarding the wafer with C ions, creating vacancies, and then annealing it, causing the vacancies to migrate and to pair into divacancies [14]. For continuous-wave ODMR measurements, we use a 975-nm laser to non-resonantly excite the electronic transitions of ensembles of defects in either a 4*H*- or 6*H*-SiC chip and an InGaAs photoreceiver to collect the near-infrared photoluminescence (PL) emitted by the defects, including their entire phonon sideband. We then use a short-terminated antenna under the chip to apply a microwave field, whose frequency (*f*) we sweep. When *f* is resonant with an electronic spin transition, the electronic spin is rotated from its optically initialized ($m_s = 0$) state towards $m_s = \pm 1$, causing changes to the PL intensity ($\Delta$PL). Although multiple defect forms are present in each of our two wafers, the frequency selectivity provided by lock-in measurements allows inequivalent defect ensembles to be measured independently [10, 14, 42].

Using low microwave-power ODMR, we observe that each electronic spin transition has a hyperfine structure (Figs. 2a-b) composed of symmetric side peaks around a central transition frequency ($f_0$). In accordance with Eq. (1), these side peaks are at frequencies $f_0 \pm A_{zz}/2$, where $A_{zz}$ is the *c*-axis projection of the hyperfine interaction between the electron spin and a nearby nucleus. The two strongest hyperfine interactions between $^{29}$Si nuclei and neutral divacancies in 4$H$-SiC are known to be at 12-13 MHz (the Si$_{IIa}$ lattice site, with 3-fold degeneracy) and 9-10 MHz (the Si$_{IIb}$ lattice site, with 6-fold degeneracy), with both hyperfine tensors nearly isotropic [5]. These sites correspond to the Si atoms nearest to the C atoms on which the neutral divacancy's electronic spin density is localized [16] (Fig. 1a).

We confirm the presence of these hyperfine interactions for neutral divacancies in 4$H$-SiC and show them also to be present for the neutral divacancies of 6$H$-SiC and the PL6 defects in 4$H$-SIC. The degeneracy of each hyperfine site is the same in every defect discussed in this paper. We use electron spin echo envelope modulation [45] to refine our measurement of the hyperfine interaction strengths [42]. Using *ab initio* density-functional theory (DFT), we then calculate the hyperfine and $D_{GS}$ constants for each form of *c*-axis-oriented neutral divacancy (Table 1). These calculations implement the plane wave and projected augmented wave method [46-48], 576- and 432-atom supercells with Γ-point sampling of the Brillouin zone, and HSE06 and PBE functionals [16, 49-52]. As has previously been done in 4$H$-SiC [5], we compare theory and experiment in order to associate each divacancy form in 6$H$-SiC with an experimentally observed spin resonance (Table 1).

We define the degree of nuclear spin polarization (*P*) as $P = (I^+ - I^-)/(I^+ + I^-)$, where $I^+$ and $I^-$ respectively represent the populations of $^{29}$Si-nuclear spins pointing ↑ and ↓ [32]. Each pair of inequivalent defect and inequivalent nuclear site has a distinct *P*. We quantify *P* by performing a global fit of the ODMR lineshape to the sum of 7 Lorentzians, one centered at $f_0$ and one pair at each of the Si$_{IIa}$, Si$_{IIb}$, and C$_a$ hyperfine resonances (Fig. 1a). We then represent $I^{\pm}$ by the fitted amplitudes of the Lorentzians at the $f_0 \pm A_{zz}/2$ side peaks (Fig. 2 and [42]). Asymmetry in the intensities of the ODMR side peaks is thus the signature of nuclear polarization.

The Boltzmann statistics of thermalized nuclear spins would require a sub-mK sample temperature (*T*) for *P* to exceed even a few percent. Indeed, at both low (*B* < 200 G) and high (*B* > 500 G) magnetic fields, we observe *P* to be nearly zero. In the 200 G < *B* < 500 G regime, however, we observe strong DNP. For PL6 defects at room temperature and *B* = 330 G, *P* reaches 99 ± 1%, an effective nuclear bath temperature of 5 μK (Fig. 2d).

Two prominent peaks can be seen in *P* as a function *B*, one centered at 300-335 G and the other at 465-490 G (Fig. 2c-d). Anticipating that level anticrossings underlie the electron-to-nuclear polarization transfer, we hypothesize that these two peaks correspond to the ESLAC and GSLAC respectively. As expected, the higher *B*-value peak in *P* corresponds precisely to $D_{GS}/(g_{GS}\mu_B)$ (Table 1) for each defect form, indicating that it is associated with the GSLAC. Due to the short (14-ns) optical lifetimes of the metastable excited states [16], though, our low-microwave-power ODMR measurements rotate spins too slowly to show ES-spin transitions.

However, high-microwave-power ODMR (Figs. 3a-c) reveals these new spin transitions. They are in the $S = 1$ electronic excited states, with $g_{ES} = 2.0$ and $D_{ES}/(g_{ES}\mu_B)$ matching precisely with the lower-$B$ peaks in $P$ (Table 1). Unlike the GS-ODMR transitions, which exhibit nonzero $\Delta$PL when microwaves and optical illumination are alternated (due to Rabi driving), they are only visible when microwaves and optical illumination are coincident [42], supporting their identification as ES resonances. Moreover, due to spin mixing in the GS, each divacancy's ES ODMR signal has a minima at its corresponding GSLAC (Fig. 3d), confirming the association between ES and GS spin transitions. Thus, peaks in $P$ (Figs. 2c-d) correspond to GSLACs and ESLACs (Figs. 3a-c).

To quantitatively understand the DNP, we simulate the optical polarization process using a recently developed model [53] of color-center-derived DNP. This model simulates the nuclear polarization while taking into account the full hyperfine tensor and the simultaneous contributions from both ESLAC- and GSLAC-derived DNP at intermediate $B$ values. In applying it, we use as many experimental parameters as possible, including $D_{GS}$, $D_{ES}$, $A_{zz}$ parameters (Table 1) and optical lifetimes [16]. The orientation of the **A** tensors are taken from our *ab initio* simulations [16], and fitting parameters represent thermally driven depolarization of the nuclear spins and the effective electron-nuclear interaction times per optical cycle. The modelled polarization (solid lines in Figs. 2c-d) and the experimental data show excellent agreement.

Our model finds that effective electron-nuclear interaction times are primarily responsible for the differences in DNP efficiencies across the different defect types. Experimentally, we use the ES spin-dephasing time ($T_{2,ES}^*$) as a proxy for the electron-nuclear interaction time, and we estimate $T_{2,ES}^*$ as $1/\pi$ times the inverse of the ES ODMR linewidth (Fig. 4a-b). As predicted, the *hh* divacancy, whose ESLAC-derived nuclear polarization is stronger than that of the $k_1k_1$ divacancy (Fig. 2c), also has a narrower ES resonance and a longer $T_{2,ES}^*$ time. Moreover, comparing ESLAC-derived DNP in SiC to that for nuclei coupled to diamond nitrogen-vacancy centers [30-35], we find that while both systems exhibit nearly ideal ESLAC-derived DNP at room temperature, the low-temperature DNP can be significantly more robust in SiC.

In diamond, both the nitrogen-vacancy center's ES-spin coherence and its off-resonantly pumped ESLAC-derived DNP rapidly decline below $T = 50$ K [34]. This diminishment is due to the deactivation of the dynamic Jahn-Teller effect, in which phonons motionally narrow pairs of ES electronic orbitals into a single coherent spin resonance [54-56]. In SiC, at $T = 5$ K, the base temperature of our cryostat, we observe both a coherent ES spin resonance (Fig. 4a-b) and strong ESLAC-derived DNP ($P = 85 \pm 5\%$ for the *hh* divacancy). As $T$ is raised, however, strong DNP persists (Fig. 4c) while $D_{ES}$ and $T_{2,ES}^*$ oscillate as a function of $T$. These behaviors suggest that while SiC Jahn-Teller effects play a role in the strong DNP, they are also complex and require further study through techniques such as pulsed ODMR in the ES [55, 57].

Our results show that strong nuclear spin polarization can be optically induced in an important industrial semiconductor. The identification of the ES-spin transitions provides an insight into the electronic structure of SiC divacancies and an understanding of the DNP process. We expect optically pumped DNP to generalize to other nuclear sites, such as the $Si_a$ and $C_a$ sites on the divacancy's symmetry axis (see Fig. 1a). Moreover, spin diffusion may also polarize nuclei that are not strongly coupled to color centers and significantly strengthen the overall nuclear spin polarization [35]. SiC nanostructures could then be used as hyperpolarized markers for magnetic resonance imaging [58, 59]. SiC is proving to have not only a key role in the power electronics and optoelectronics industries but also in the fields of spintronics, sensing, and quantum information.

**Acknowledgments**

The authors thank Viatcheslav V. Dobrovitski, Bob B. Buckley, F. Joseph Heremans, and Charles de las Casas for helpful conversations. This work is supported by the Air Force Office of Scientific Research, the National Science Foundation, the Material Research Science and Engineering Center, the Knut & Alice Wallenberg Fund, the Lendület program of the Hungarian Academy of Sciences, and the National Supercomputer Center in Sweden.

## Figures

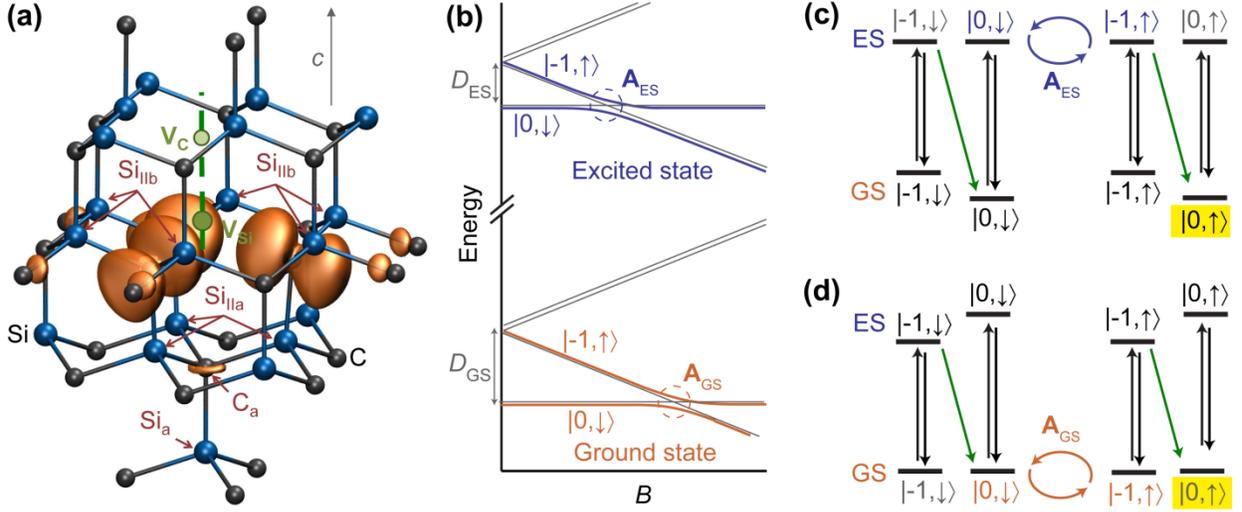

**Fig. 1.** **(a)** An illustration of the $k_1k_1$ divacancy (green circles) in $6H$-SiC. The calculated spin density is represented by orange-lobe isosurfaces and primarily localized to the dangling bonds of the Si vacancy's nearest C atoms. We measure the DNP of $^{29}$Si nuclei at the $Si_{IIa}$ and $Si_{IIb}$ sites. **(b)** Evolution of the ES and GS spin-sublevel energies with $B$, showing mixing at the hyperfine-mediated ESLAC and GSLAC. The states drawn in gray do not mix. Both $D_{ES}$ and $D_{GS}$ are positive [42]. **(c)** For ESLAC-derived DNP, a hyperfine interaction in the ES causes $|0,\downarrow\rangle$ to partially evolve into $|-1,\uparrow\rangle$ every optical cycle. Together with the electron spin-spin polarization provided by the intersystem crossing (green arrows), this interaction causes optical cycling (black arrows) to polarize arbitrary $|m_S, m_I\rangle$ states into $|0,\uparrow\rangle$ (highlighted). The $m_s = +1$ electronic spin states do not participate in this process. **(d)** For GSLAC-derived DNP, the mechanism is the same as in (c), but the relevant hyperfine interaction is in the GS.

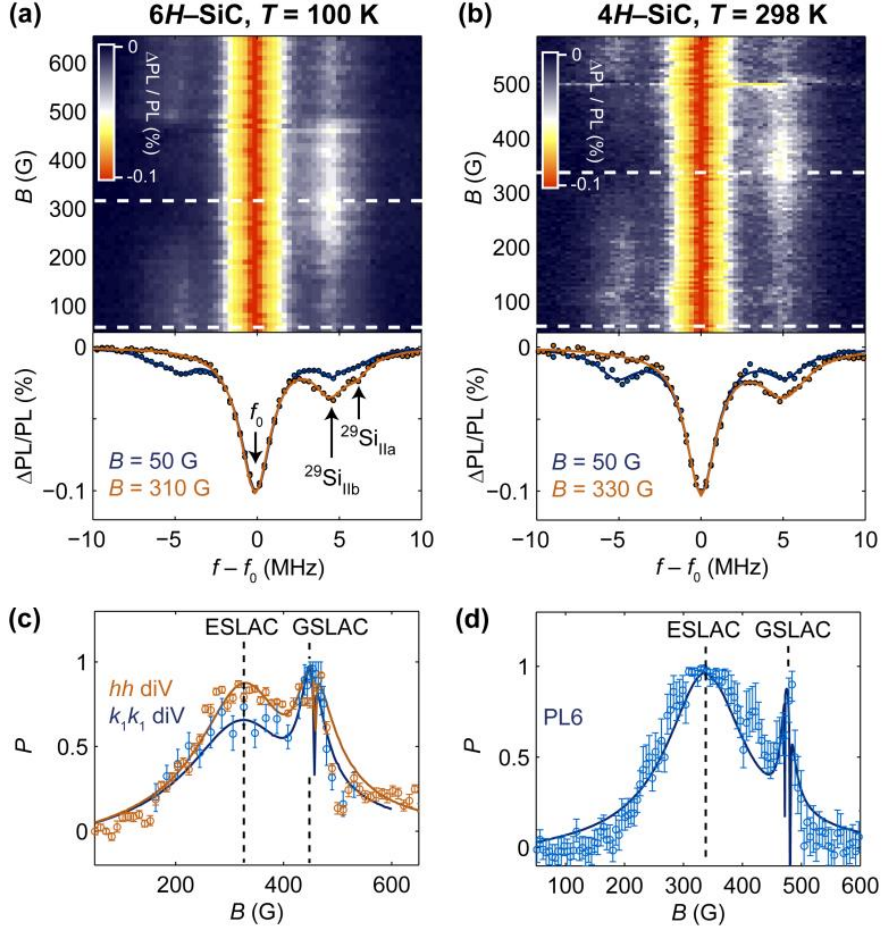

**Fig. 2. (a)** Upper: Low-microwave-power ODMR spectrum of the $m_s = 0$ to $m_s = 1$ spin transition of the *hh* divacancy in 6*H*-SiC at $T = 100$ K. As $B$ varies from 50 G to 650 G, $f_0$ varies from 1.5 GHz to 3.2 GHz, and strong $P$ is seen at intermediate $B$ values. Lower: Line cuts at the dashed white lines ($B = 50$ G and $B = 310$ G). The $^{29}$Si$_{IIa}$ and $^{29}$Si$_{IIb}$ nuclei are unpolarized at $B = 50$ G and nearly completely polarized at the ESLAC ($B = 310$ G). The continuous lines are fits to sums of Lorentzians [42]. **(b)** Upper: Low-power ODMR spectrum of the $m_s = 0$ to $m_s = 1$ spin transition of the PL6 defects in 4*H*-SiC at $T = 298$ K. Lower: Line cuts at $B = 50$ and at $B = 330$ G, which is the PL6 ESLAC. **(c)** $^{29}$Si$_{IIb}$ nuclear polarization ($P$) for nuclei coupled to *hh* and $k_1k_1$ divacancies in 6*H*-SiC at $T = 100$ K, exhibiting peaks in $P$ at the ESLAC and GSLAC. *hh* divacancies have stronger ESLAC-related DNP than $k_1k_1$ divacancies. We plot $P$ at the $^{29}$Si$_{IIa}$ sites in [42]. Due to spectral overlap with the stronger *hh* divacancies, $P$ for nuclei coupled to $k_2k_2$ divacancies in 6*H*-SiC could not be accurately measured. The error bars are single-σ confidence intervals set by the fits. The continuous lines are $P$ values simulated from our theoretical model. **(d)** $^{29}$Si$_{IIb}$ nuclear polarization for nuclei coupled to PL6 defects in 4*H*-SiC at $T = 298$ K (experiment and theory). The origin of the peak in $P$ at 400 G is unknown.

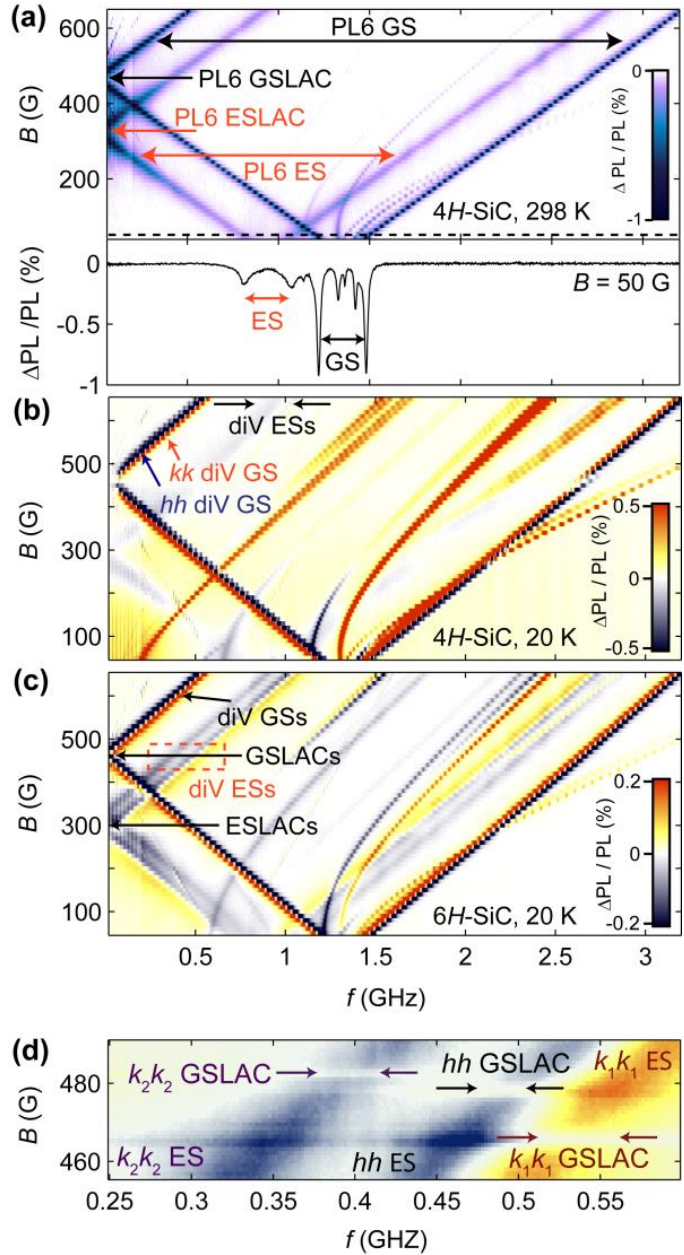

**Fig. 3. (a)** Upper: High-power ODMR spectrum of PL6 at $T = 298$ K. Lower: Line-cut of the ODMR spectrum at $B = 50$ G (the dashed line). **(b)** High-power ODMR spectrum of the neutral divacancies in 4$H$-SiC at $T = 20$ K. **(c)** High-power ODMR spectrum of the neutral divacancies in 6$H$-SiC at $T = 20$ K. **(d)** Zoom-in of the red dashed rectangle drawn in (c). Due to spin mixing in the GS, each divacancy's ES ODMR signal has a minima at its corresponding GSLAC.

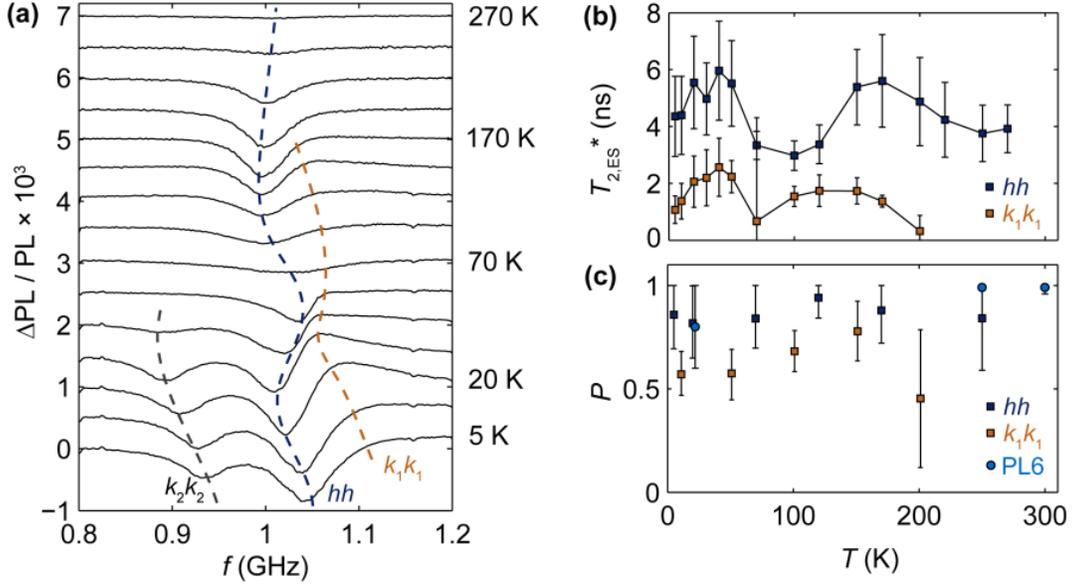

**Fig. 4. (a)** Temperature dependence of the high-power ES ODMR for the three c-axis-oriented neutral divacancies in $6H$-SiC at $B = 650$ G. The ES ODMR curves are sequentially offset from $\Delta PL = 0$, for clarity, and the dashed lines are guides to the eye that follow the ES-spin resonances. **(b)** Power-broadened $T_{2,ES}^*$ for divacancies in $6H$-SiC, calculated by fitting the curves in (a) to a sum of three Lorentzians and taking $T_{2,ES}^*$ to be $1/\pi$ times the inverse of the linewidths. The error bars are single-$\sigma$ uncertainties set by the fits. **(c)** Temperature dependence of $P$ at the $Si_{IIb}$ site of divacancies in $6H$-SiC, and for PL6 in $4H$-SiC, when $B$ is tuned to the ESLAC (270-330 G). The single-$\sigma$ error bars derive from fits to ODMR spectra [42].

| Defect | ZPL (eV) | Sign of ΔPL | $D_{ES}$ (GHz) | $D_{GS}$ (GHz) | $A_{zz}$ Si$_{IIa}$ (MHz) | $A_{zz}$ Si$_{IIb}$ (MHz) | $D_{GS}$ (GHz) | $A_{zz}$ Si$_{IIa}$ (MHz) | $A_{zz}$ Si$_{IIb}$ (MHz) |
|---|---|---|---|---|---|---|---|---|---|
| | | | 4H-SiC *experiment* | | | | 4H-SiC *calculation* | | |
| **hh diV** | 1.095 | + | 0.84 | 1.336 | 12.3 | 9.2 | 1.358 | 11.6 | 9.3 |
| **kk diV** | 1.096 | − | 0.78 | 1.305 | 13.2 | 10.0 | 1.320 | 12.4 | 10.2 |
| **PL6** | 1.194 | − | 0.94 | 1.365 | 12.5 | 9.6 | -- | -- | -- |
| | | | 6H-SiC *experiment* | | | | 6H-SiC *calculation* | | |
| **hh diV** | 1.092 | + | 0.85 | 1.334 | 12.5 | 9.2 | 1.350 | 11.8 | 9.6 |
| **$k_1k_1$ diV** | 1.088 | − | 0.75 | 1.300 | 12.7 | 10.0 | 1.300 | 12.7 | 10.5 |
| **$k_2k_2$ diV** | 1.134 | − | 0.95 | 1.347 | 13.3 | 9.2 | 1.380 | 11.8 | 9.7 |

**Table 1.** The zero-field parameters, the zero-phonon line (ZPL) optical transition energies, and the ground-state $A_{zz}$ values for $^{29}$Si nuclei coupling to the PL6 defect and *c*-axis-oriented neutral divacancies in 4H- and 6H-SiC. Both $D_{GS}$ and $D_{ES}$ are positive [42]. All parameters are at $T = 20$ K, except for $D_{ES}$ of PL6, where the room-temperature value is provided. The calculations of $D_{GS}$ and $A_{zz}$ are at $T = 0$ K, using the method in Ref. [16]. We match the divacancy forms in 6H-SiC with their corresponding spin transitions by comparing the experimentally determined and calculated $D_{GS}$ parameters.

**Supplemental Material for "Optical polarization of nuclear spins in silicon carbide"**

Abram L. Falk[1,2], Paul V. Klimov[1,3], Viktor Ivády[4,5], Krisztián Szász[4,6], David J. Christle[1,3], William F. Koehl[1], Ádám Gali[4,7], and David D. Awschalom[1,*]

1. Institute for Molecular Engineering, University of Chicago, Chicago, IL 60637, USA
2. IBM T. J. Watson Research Center, 1101 Kitchawan Rd., Yorktown Heights, NY 10598, USA
3. Center for Spintronics and Quantum Computation, University of California, Santa Barbara, Santa Barbara, CA 93106, USA
4. Institute for Solid State Physics and Optics, Wigner Research Centre for Physics, Hungarian Academy of Sciences, Budapest, Hungary
5. Department of Physics, Chemistry, and Biology, Linköping University, Sweden
6. Institute of Physics, Loránd Eötvös University, Hungary
7. Department of Atomic Physics, Budapest University of Technology and Economics, Budapest, Hungary
* Email: awsch@uchicago.edu

PACS numbers: 76.70.Fz, 76.30.Mi, 71.55.-i, 61.72.jn, 42.50.Ex, 85.75.-d

**Contents:**

1. Experimental techniques: sample preparation and optically detected magnetic resonance

2. Concentration of polarized nuclei

3. Electron-spin echo envelope modulation measurements of hyperfine energies

4. Sign of hyperfine and zero-field splitting parameters

5. Methods for fitting nuclear polarization from optically detected magnetic resonance

6. Further characterization of the PL6 excited state

**1. Experimental techniques: Sample preparation and optically detected magnetic resonance (ODMR)**

In addition to the following summary, our sample-preparation methods and experimental setup are described in detail in Refs. [S1, 2]. As described in the main text, the 4$H$-SiC samples are high-purity semi-insulating wafers purchased from Cree, Inc (part number: W4TRD0R-0200). Since they contain "off-the-shelf" neutral divacancies and PL6 defects, we dice them into chips and measure them without any further sample preparation. We purchase 6$H$-SiC wafers from II-VI, Inc. In order to incorporate neutral divacancies into the 6$H$-SiC samples, we first have them bombarded with $^{12}$C ions at an energy of 200 keV with doses ranging from $10^{11} – 10^{13}$ cm$^{-2}$ (Cutting Edge Ions, Inc.), which generates crystal vacancies. Using James Ziegler's Stopping Range of Ions in Matter software (http://www.srim.org), we estimate that the vacancies from the $^{12}$C ions are localized to the top 500 nm of our SiC chips. We then anneal the samples at 950 °C for 30 minutes in order for the vacancies to migrate and form vacancy complexes, including divacancies.

For ODMR measurements, we use a 300 mW, 1.27 eV (975 nm) diode laser, purchased from Thorlabs, Inc. 60 mW reaches the sample. For measurements that required the laser to be gated (which are in this Supplemental Material only), we use an acousto-optical modulator. We focus the laser excitation onto the sample using a near-infrared coated 50× Olympus objective and collect the photoluminescence (PL) using that same objective. We then focus the collected PL onto an InGaAs photoreceiver, which was purchased from FEMTO, a German electronics manufacturer. The SiC samples are 3-4 mm chips attached to coplanar microwave striplines with rubber cement. In turn, the microwave stripline is soldered to a copper cold finger, which is cooled by a Janis flow cryostat.

We use lock-in techniques to take all of the data in this paper. For the continuous-wave ODMR measurements, we leave the excitation laser on while gating the microwave signal on and off at 1-5 kHz. The change in PL ($\Delta$PL) when the microwave is on is the ODMR signal. We plot the ODMR signal as a normalized ratio: $\Delta$PL/PL. However, due to background PL from the sample (i.e. PL unrelated to the defect of interest), this fractional change in PL is much lower than the intrinsic fractional change in PL of the neutral divacancies and PL6 (i.e. the fractional change when there is no background PL). Using single-spin measurements, we have found the intrinsic $\Delta$PL/PL of neutral divacancies to be roughly 15% [S3] at 20 K. We have also used spectrally resolved ensemble measurements to estimate that the intrinsic $\Delta$PL/PL of PL6 defects is 35% [S4].

When we sweep the frequency ($f$) outputted by our microwave signal generators, we observe oscillations in the microwave power reaching the sample, largely the result of cable reflections. In order to stabilize the microwave power that reaches the sample, we measure the microwave power with a Schottky diode and feed back the signal-generator power to stabilize the voltage on the Schottky diode. Due to the dispersion of the Schottky diode, this technique still allows some fluctuations in microwave power (in particular, higher powers at lower frequencies), but it significantly flattens the power-response curve.

Our measurements use both "low microwave powers" (Fig. 2 and Fig. 4c-d) and "high microwave powers" (Fig. 3 and Fig. 4a). The low microwave power is used to mitigate power broadening of the spin transitions and obscure the hyperfine structure of the ODMR lines. For these measurements, 2 mW of microwave power is sent into to the cryostat, corresponding to Rabi frequencies of ~500 kHz. The high power-microwave measurements are used to rotate electron spins fast enough to observe ES ODMR. For these measurements, 1 W of microwave power is sent to the cryostat, corresponding to ~10 MHz Rabi frequencies.

## 2. Concentration of polarized nuclei

Using double electron-electron resonance, we have previously found that our 6*H*-SiC samples that were bombarded with the highest ($10^{13}$ cm$^{-2}$) dose of $^{12}$C ions (and then annealed) have a roughly $10^{16}$ cm$^{-3}$ density per *c*-axis-divacancy species [S2]. Using the 3 total species of c-axis-oriented neutral divacancies in 6*H*-SiC, the 3 (6) possible $^{29}$Si sites per divacancy in the Si$_{IIa}$ (Si$_{IIb}$) hyperfine sites, and the 4.7% natural abundance of SiC, these samples then have $^{29}$Si nuclei at a concentration of $10^{16}$ cm$^{-3}$. Due to the diffusion of nuclear polarization, this density should be taken as a lower bound and may be a significant underestimate. Measuring 6*H*-SiC samples with a $^{12}$C ion dose of both $10^{12}$ and $10^{13}$ cm$^{-2}$, we have observed no difference in dynamic nuclear polarization (DNP) as a function of $^{12}$C ion dose (and thus divacancy density) in these two types of substrates.

We can estimate the density of neutral divacancies and PL6 defects in our 4*H*-SiC sample by comparing the intensity of the zero-phonon-line (ZPL) peaks in its PL spectrum to those in 6*H*-SiC, measured with identical optics under identical experimental conditions. We find that the integrated intensity of each *c*-axis-oriented-divacancy ZPL in 4*H*-SiC is roughly ¼ as intense as that in the 6*H*-SiC sample bombarded with the $10^{13}$ cm$^{-2}$ dose of C ions. However, due to the ion bombardment procedure that generated the neutral divacancies in our 6*H*-SiC samples, the divacancies in 6H-SiC are localized to the top 500 nm of the chip, whereas the as-grown neutral divacancies in 4H-SiC are distributed throughout the full 500 μm chip. The neutral divacancies in our 4*H*-SiC samples thus have a density of roughly $3 \times 10^{12}$ cm$^{-3}$ per defect form.

For the PL6 defects in our 4*H*-SiC substrate, we do not want to make an assumption about the depth distribution of the defects. Thus, we characterize the areal concentration of these defects rather than the volumetric concentration. Each c-axis-divacancy form in 6*H*-SiC has an areal density of ~$5 \times 10^{11}$ cm$^{-2}$ in the plane of the wafer. The integrated ZPL of PL6 is 25 times less intense than that of the 6*H*-SiC divacancies. The PL6 defects thus have an areal density of $2 \times 10^{10}$ cm$^{-2}$ in the plane of the wafer.

## 3. Electron-spin echo envelope modulation (ESEEM) measurements of hyperfine energies

Hyperfine spectra can in principle be measured by ODMR, but these measurements are limited by the inhomogeneously broadened spin transition linewidths (~$1/T_2^*$), which range from 1- 3 MHz in our samples. Since this is of the same order as the differences in the various hyperfine lines, accurately resolving them is challenging. To surpass this limitation, we extract the hyperfine coupling parameters via ESEEM.

Hyperfine spectra extracted via ESEEM are not limited by the electron's $T_2^*$ (as with ODMR), but rather their much longer homogeneous coherence time ($T_2$) (assuming that nuclear spin relaxation times are even longer than that). We use "two-pulse ESEEM", which is based on simple Hahn-echo coherence ($C(t_\text{free})$) [S5]. "Two-pulse ESEEM" is really three pulses in an ODMR measurement (as opposed to tradition electron-spin resonance), due to the final $\pi/2$ pulse which projects the spin coherence to the measurement axis. These data can be generally fit to a product of echo envelope modulation functions ($E_j(t_\text{free})$) times an overall decoherence function $D(t_\text{free})$:

$$C(t_\text{free}) = D(t_\text{free}) \prod_j E_j(t_\text{free}).$$

(S1)

Each $E_j(t_\text{free})$ is given by:

$$E_j(t_\text{free}) = 1 - 2K_{\pm 1} \sin^2(\pi \nu_0 t_\text{free}) \sin^2(\pi \nu_{\pm 1} t_\text{free}),$$

(S2)

where the $\nu_{0,\pm 1}$ correspond to the nuclear spin splittings in the electronic $m_s = 0$ and $m_s = \pm 1$ branches respectively, and $K_{\pm 1}$ are the modulation depth parameters, which characterize the amplitude of the ESEEM oscillations [S5]. The $\nu_0$-frequency oscillation is a relatively slow oscillation at the frequency corresponding to the nuclear Zeeman interaction. The $\nu_{\pm 1}$ oscillations correspond to the sum of the nuclear Zeeman interaction and the hyperfine interaction. However, for the $^{29}\text{Si}_\text{IIa}$ and $^{29}\text{Si}_\text{IIb}$ hyperfine interactions that we study here, the nuclear Zeeman effect at $B = 40$ G is much smaller than the hyperfine interactions: since it will shift $\nu_{0,\pm 1}$ by only 40 kHz, and the uncertainty of our measurement is 100 kHz, we neglect it. Thus, in accordance with Eq. S2, the $c$-axis projection of the hyperfine tensor can be simply read off as half the frequencies of peaks of the fast Fourier transform (FFT) of $C(t_\text{free})$ (Fig. S1).

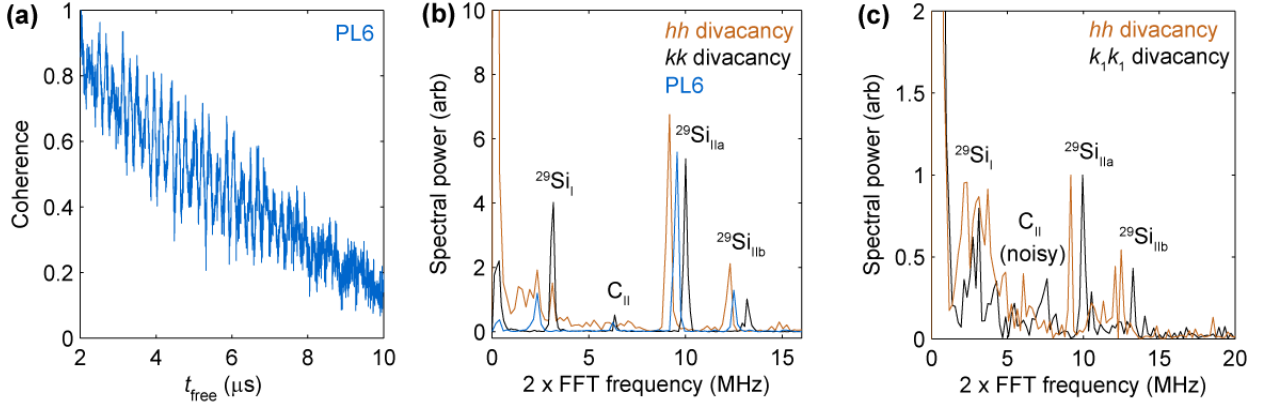

**Figure S1.** (a) Hahn-echo coherence of PL6 at room temperature as a function of $t_{\text{free}}$, at an external field of $B = 40$ G. The frequency of ESEEM oscillations correspond to hyperfine interaction strengths. The apparently overall decay in coherence is due to Larmor precession of $^{29}$Si in the sample: the coherence will refocus at later times. (b). The FFT of the Hahn-echo coherence (ESEEM oscillations) of the 4$H$-SiC divacancies and PL6 defects at $T = 20$ K shows resolvable 4 sites for hyperfine-coupled nuclear spins. The hyperfine peaks are labeled according to the convention in Ref. [S6], except for the $C_{II}$ peak at 6.3 MHz, which was not seen there. (c) The FFT of Hahn-echo coherence (ESEEM oscillations) from the $hh$ and $k_1k_1$ neutral divacancies in 6$H$-SiC. We do not plot the data from the $k_2k_2$ divacancies, because it exhibits a lot of spurious features due to interference from the $hh$ divacancy. Nonetheless, we can (more roughly) estimate the hyperfine strengths of $^{29}$Si in the Si$_{IIa-b}$ lattice sites for this defect. The hyperfine interaction strengths are summarized in Table 1 of the main text.

### 4. Sign of hyperfine and zero-field splitting parameters

To establish the sign of the nuclear polarization, we first infer from the sign of the polarized electron spins' magnetization, previously measured using double electron-electron resonance [S2], that the electron spins bound to neutral divacancies and PL6 defects are all-optically initialized into the $m_s = 0$ spin sublevel. Then, we note that first-principles calculations have shown that the sign of $A_{z,\text{GS}}$ is positive. $D_{\text{GS}}$ is also known to be positive from both electron spin-resonance measurements [S6] and first-principles calculations [S4]. These two observations, coupled with Eq. (1) and the observation that DNP populates the higher energy hyperfine transition, indicate that the $^{29}$Si nuclei are optically polarized into the $\uparrow$ state. In order for mixing between electronic and nuclear spin states in the ES to conserve angular momentum, $|0, \downarrow\rangle$ must couple to $|-1, \uparrow\rangle$ in either the GS or ES. In turn, in order for the $m_s = -1$ and $m_s = 0$ electronic to nearly intersect and allow this mixing to occur, the sign of $D_{\text{ES}}$ must be positive.

### 5. Methods for fitting nuclear polarization from optically detected magnetic resonance

The degree of nuclear-spin polarization ($P$) is determined by the relative amplitudes of their continuous-wave ODMR peaks. Peaks associated with different hyperfine splittings, which in our case differ by only a few MHz, will overlap somewhat due to a finite T2* ($< 1$ μs), which can make the determination of their exact amplitudes non-trivial. Moreover, the high $P$ values near the GSLAC and ESLAC for our systems (>90%) and finite measurement noise present in the recorded data mean that conventional least squares/maximum likelihood techniques will sometimes yield estimates of $P$ that exceed 100%. To accurately infer the $P$ values, we adopt a Bayesian approach that is analogous to a global nonlinear least-squares fit. In particular, the most general model for the signal $y_{jk}(f_{jk})$ measured in each sweep of frequency $f_{jk}$ includes three pairs of Lorentzian lineshapes corresponding to hyperfine splittings about a central Lorentzian resonance,

$$y_{jk}(f_{jk}) = y_{0,j} + a_0 \frac{g_0^2}{g_0^2 + (f_{jk} - \delta_j)^2} + \sum_{i=1}^{3} \left( a_{1ij} \frac{g_i^2}{g_i^2 + (f_{jk} - \vartheta_i - \delta_j)^2} + a_{2ij} \frac{g_i^2}{g_i^2 + (f_{jk} + \vartheta_i - \delta_j)^2} \right),$$

(S3)

where $j$ indexes the sweep number, $k$ indexes the individual data points within the sweep, $y_{0,j}$ is a small constant signal offset, $\vartheta_i$ are the hyperfine splittings, $\delta_j$ is a small frequency offset, the $a$ terms are the Lorentzian amplitudes, and the $g$ terms are the Lorentzian scale parameters. The amplitudes $a_{1ij}$ and $a_{2ij}$ are, in particular, the right and left hyperfine resonance for the $i$'th hyperfine resonance pair, respectively. Before fitting the model, we normalize and center the sweeps by computing the maximum signal data point for each sweep, about which the sweep is then centered, and divide the dataset by this maximum value in order to normalize it. Because of measurement noise, this procedure produces a slight jitter in the central frequency and the inferred amplitude of the central peak, which we compensate for with the small $\delta_j$ frequency offsets for the Lorentzians above, and a moderately strong prior: $a_0 \sim Normal(1.00, 0.05^2)$.

Practically speaking, this procedure allows for small deviations from unity for the central peak amplitude. In the model, the $g_i$ and $f_i$ terms are shared between sweeps to enhance the precision to which we can determine the other parameters of the model. We then relate these ideal $y_{jk}$ to the observed data through the standard least squares likelihood up to an overall unimportant normalization,

$$\log\left(p(y_{jk}|\theta)\right) = \log\left(L(y_{jk})\right) = -\frac{(y_j - d_{jk})^2}{2\sigma_j^2} - \frac{1}{\sigma_j},$$

(S4)

where $\sigma_j$ is a free parameter for the standard deviation of our measurement noise after normalization. This figure is typically about 1% for most sweeps.

We sample from the posterior distribution using a Markov Chain Monte Carlo technique [S7] and use the samples to compute the polarization using the relation $P_{ij} = (a_{1ij} - a_{2ij})/(a_{1ij} + a_{2ij})$. Since the Bayesian approach bounds the amplitudes to be greater than or equal to 0, the polarization is always in the physical region between -1 and 1. Transforming the samples from the posterior gives the marginal distribution of $P_{ij}$, which we use to compute the mode and the highest posterior density intervals (which are also only within the physical region, for the same reason) that are reported in the main text.

Since the PL6 polarization data is taken at room temperature, larger $T_2^*$ values makes calculating $P$ more challenging. When analyzing these data, we fix the outermost hyperfine resonances at their values extracted from a Fourier transform of a separate electron spin echo envelope modulation (ESEEM) measurement. We also make the approximation that the polarizations of the $Si_{IIa}$ and $Si_{IIb}$ resonances are equal, which means that the corresponding amplitudes have a constant (free-parameter) ratio.

## 6. Further characterization of the PL6 excited state

Here, we provide two additional measurements of the PL6 excited state (Fig. S2a). First, in order to prove that the GS and ES spin transitions that we see in ODMR are associated with each other, we observe that the ES-ODMR line has a minimum in ΔPL precisely at the PL6 GSLAC (Fig. S2b). The relationship between Figs. S2a and S2b is closely analogous to that of Fig. 3c and Fig. 3d in the main text. Second, we show that, when we try to measure the ES with pulsed ODMR, the ES can only be seen when laser and microwave excitations coincide (Fig. S2c). In contrast, GS ODMR can be seen as Rabi oscillations both when the laser and microwave excitations coincide, and also when they alternate.

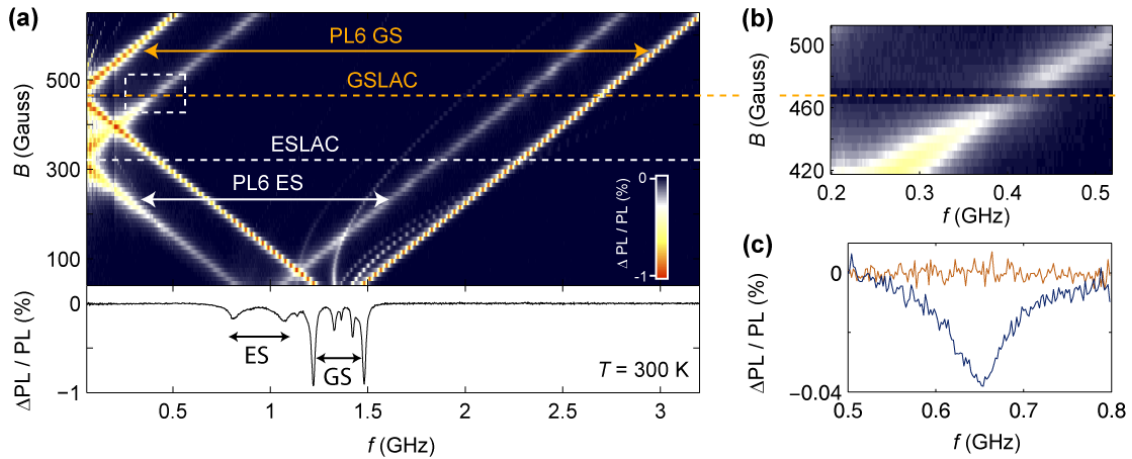

**Figure S2. (a)** High-power ODMR spectrum of PL6 as a function of $f$ and $B$ at room temperature. Lower: Line-cut of the ODMR spectrum at $B = 50$ G. This panel is a reproduction of Fig. 3(a) from the main text. **(b)** At the PL6 GSLAC, the PL6 ES loses its ODMR visibility, thereby associating the ES and GS spin transitions seen in ODMR. **(c)** ODMR of the PL6 excited state at $B = 100$ G, with both the microwave and excitation light pulsed at 1 kHz with a 50% duty cycle. When the excitation light and microwave pulses coincide, an excited state ODMR signal is seen (blue curve). When the light and microwaves alternate, there is no excited-state ODMR (orange curve).